\documentclass[aps,prl,twocolumn,superscriptaddress,longbibliography]{revtex4-2}
\usepackage{amsfonts}
\usepackage{amsmath}
\usepackage{mathtools}
\usepackage{longtable}
\usepackage{amssymb}
\usepackage{graphicx}
\usepackage{bm}
\usepackage{color}
\usepackage{mathrsfs}
\usepackage[colorlinks,bookmarks=true,citecolor=blue,linkcolor=red,urlcolor=blue]{hyperref}
\usepackage{appendix}
\usepackage{float}
\usepackage{booktabs}
\usepackage{esint}
\usepackage[export]{adjustbox}

\newcommand{\RNum}[1]{\uppercase\expandafter{\romannumeral #1\relax}}

\begin{document}
\title{Superdiffusive Transport in Quasi-Particle Dephasing Models}

\author{Yu-Peng Wang}
\affiliation{Beijing National Laboratory for Condensed Matter Physics and Institute of Physics, Chinese Academy of Sciences, Beijing 100190, China}
\affiliation{University of Chinese Academy of Sciences, Beijing 100049, China}
\author{Chen Fang}
\affiliation{Beijing National Laboratory for Condensed Matter Physics and Institute of Physics, Chinese Academy of Sciences, Beijing 100190, China}
\affiliation{Songshan Lake Materials Laboratory, Dongguan, Guangdong 523808, China}
\affiliation{Kavli Institute for Theoretical Sciences, Chinese Academy of Sciences, Beijing 100190, China}
\author{Jie Ren}
\email{jieren@iphy.ac.cn}
\affiliation{Beijing National Laboratory for Condensed Matter Physics and Institute of Physics, Chinese Academy of Sciences, Beijing 100190, China}
\affiliation{University of Chinese Academy of Sciences, Beijing 100049, China}

\begin{abstract}
Investigating the behavior of noninteracting fermions subjected to local dephasing, we reveal that quasi-particle dephasing can induce superdiffusive transport. 
This superdiffusion arises from nodal points within the momentum distribution of local dephasing quasi-particles, leading to asymptotic long-lived modes. 
By studying the dynamics of the Wigner function, we rigorously elucidate how the dynamics of these enduring modes give rise to L\'evy walk processes, a renowned mechanism underlying superdiffusion phenomena. 
Our research demonstrates the controllability of dynamical scaling exponents by selecting quasi-particles and extends its applicability to higher dimensions, underlining the pervasive nature of superdiffusion in dephasing models.
\end{abstract}

\maketitle

\textit{Introduction.---}
Transport properties of particles, energy, and information in nonequilibrium quantum many-body systems have garnered significant attention \cite{transport1, transport2, transport3, heat1, heat2, heat3, info1, info2, info3, info4, info5, info6}. 
The emergence of anomalous transport, which deviates from the classical diffusion characterized by linear growth in mean square displacement over time, challenges established principles in quantum many-body dynamics. 
This anomaly includes superdiffusion \cite{superdiffusion1} and subdiffusion \cite{subdiffusion1}, where particle spreading occurs faster or slower than classical expectations.

Notably, the one-dimensional Heisenberg model has exhibited superdiffusion \cite{znidaric11, prosen17, prosen19, moore18, moore21, ilievski19, vasseur19, vasseur20} with a dynamical exponent of $z=3/2$ and a scaling function within the Kardar-Parisi-Zhang (KPZ) universality class \cite{kpz1,kpz2,kpz3}. 
This superdiffusive behavior extends to a broader class of integrable models with non-Abelian symmetries \cite{symm1, symm2, symm3, symm4, symm5, symm6}, with transitions to diffusive behavior observed when integrability or symmetry is perturbed \cite{diffusive1, diffusive2, diffusive3, diffusive4}. 
Superdiffusion has also been identified in systems with long-range interactions \cite{longrange1, longrange2, longrange3, longrange4, longrange5, longrange6} and short-range interacting systems subject to quasiperiodic potentials \cite{AAH1, AAH2}, albeit with some controversies \cite{AAH3}.

So far, the study of superdiffusion mainly focuses on closed systems since it is generally believed that coupling a system to the environment results in bulk dissipation, leading to diffusion.
A well-studied example is the free fermion chain subject to local particle dephasing \cite{dephase1,dephase2,dephase3,dephase4,dephase5,wigner1,wigner2}. 
In this context, dephasing introduces finite lifetimes to the original free modes, resulting in a mean free path beyond which particle motion resembles a Gaussian random walk.

In contrast, our study identifies a superdiffusive transport in noninteracting fermion systems by generalizing the onsite dephasing to ``quasi-particle" dephasing. 
The ``quasi-particles" are defined as superpositions of fermions near position $x$: $\hat d_x = \sum_a d_a \hat c_{x+a}$, where the vector ${d_a}$ is assumed to be local near the origin. 
The momentum distribution characterizing these quasi-particles is $d_k \equiv \sum_a d_a e^{ika}$.
Remarkably, our investigation unveils a direct link between the nodal structure of $d_k$ and the occurrence of superdiffusion:
\begin{enumerate}
	\item When $d_k$ possesses a nodal point at generic momentum $k_o$ (with nonzero velocity $v_{k_o} \ne 0$) characterized by $|d_k| \sim (k-k_o)^{-n}$, particle transport exhibits a ballistic front, and the dynamical scaling exponent is given by $z_n=(2n+1)/(2n)$;
	\item In cases where $d_k$ features a higher-order nodal point at zero-velocity point $k_o$, described as $|d_k| \sim (k-k_o)^{-n}$ where $n \geq 2$, the particle transport exhibit a superdiffusive front, and the dynamical exponent is $z_n = (2n+1)/(2n-1)$.
\end{enumerate}

We demonstrate the superdiffusion in dephasing models by analyzing the dynamics of the Wigner function \cite{wigner1,wigner2}.
This approach develops a hydrodynamics description effective in capturing transport behavior across extended temporal and spatial scales.
The hydrodynamics framework translates the many-body transport problem into a single-particle random walk process.
In this context, the presence of nodal points signifies the existence of long-lived modes with diverging mean free paths.
The probabilistic distribution characterizing these mean free paths exhibits a heavy-tailed nature, a hallmark of the L\'{e}vy walk \cite{levy}, a well-established model of superdiffusive processes.

Significantly, the dynamical exponent of the charge transport is intricately linked to the nodal structure of the dephasing quasi-particle. 
In instances where certain symmetries are present, the presence of nodal points becomes generic, leading to a robust manifestation of superdiffusion characterized by exact dynamical exponents.
Besides, fine-tuning the nodal structure of quasi-particles enables the systematic generation of a spectrum of dynamical exponents. 
Our analytical approach extends naturally to higher-dimensional systems, underscoring the universality of our findings in the context of dephasing models.

\begin{figure*}
	\centering
	\includegraphics[width=\linewidth]{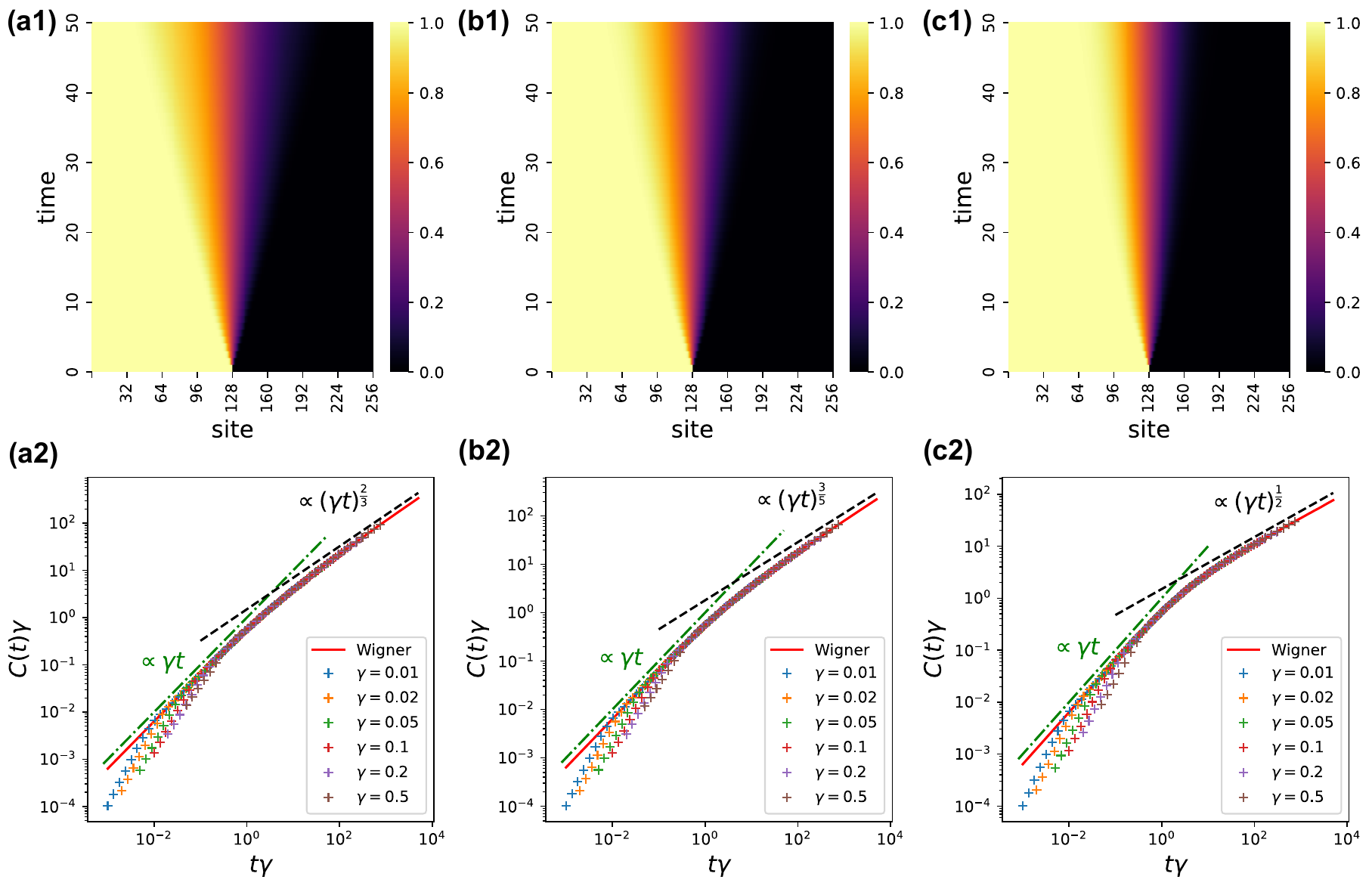}
	\caption{Numerical simulations of the particle transport of the dephasing Lindblad equation (\ref{eq:lindbladian}), with $\hat{d}_x$ defined according to Eq.~(\ref{eq:generic-nodal}). Subplots are presented for different parameters, namely (a1)(a2) for $a=\sqrt{2}$, (b1)(b2) for $a=2$, and (c1)(c2) for $a=3$. Subplots (a1)(b1)(c1) display the density evolution of systems with different dephasing quasi-particles. The system sizes are fixed to $L=256$, and the dephasing strength is $\gamma=0.5$. The dynamics are initiated from the domain-wall state $|\psi_o\rangle=|1,\dots,1,0,\dots,0\rangle$, and feature either (a1) a ballistic wavefront, (b1) a superdiffusive wavefront, or (c1) a diffusive wavefront. Subplots (a2)(b2)(c2) shows the charge transport $C(t)$ for different dephasing quasi-particles. The system size for this simulation is fixed to $L=3000$. For exact Lindbladian simulation, in the short time regime ($t<1$), the transport behaviors deviate from the hydrodynamics results, following a $C(t)\propto t^{2}$ scaling. After $t > 1$, the Lindbiadian result approaches the hydrodynamics, which exhibits a crossover from ballistic ($z=1$) to (a2) superdiffusive with dynamical exponent $z=3/2$, (b2) superdiffusive with dynamical exponent $z=5/3$, and (c2) diffusive with dynamical exponent $z=2$.}
	\label{fig:fig1}
\end{figure*}

\textit{Dephasing models.}---
Consider the dynamics of a dephasing model governed by the Lindbladian:
\begin{equation}\label{eq:lindbladian}
	\partial_t \hat \rho = -i[\hat H,\hat \rho]-\frac{\gamma}{2}\sum_x [\hat L_x,[\hat L_x,\hat\rho]],
\end{equation}
where the Hamiltonian $\hat H$ represents a basic noninteracting fermion chain, given by $\hat H = \sum_i (\hat c_i^\dagger \hat c_{i+1} + \hat c^\dagger_{i+1} \hat c_i)$, characterized by a group velocity $v_k=2\sin k$. 
The jump operator $\hat L_x = \hat d_x^\dagger \hat d_x$ captures the dephasing process affecting the quasi-particles $\hat d_x$. 
In the previous studies, particularly in the context of both monitored \cite{wigner1} and open systems \cite{dephase2}, the case where $\hat d_x = \hat c_x$ has been well-explored, resulting in a clear demonstration of diffusive particle transport.

We first focus on the scenario involving quasi-particles with time-reversal and spatial-reflection symmetry. 
Specifically, we examine a scenario involving a three-site quasi-particle configuration:
\begin{equation}\label{eq:generic-nodal}
	\hat d_x = \frac{1}{\sqrt{2+a^2}}(\hat c_{x-1} - a \hat c_{x} + \hat c_{x+1}),
\end{equation}
where $a$ is a real parameter. 
The corresponding momentum distribution is $d_k = (2 \cos k - a) / \sqrt{2+a^2}$. 
In the range $-2<a<2$, $d_k$ exhibits two nodal points, $k_\pm = \pm \arccos(a/2)$, around which $d_k$ is linearly dispersed.
Upon reaching $a = \pm 2$, these two nodal points merge into a higher-order nodal point at $k=0$ or $k=\pi$ with quadratic dispersion: $d_k \propto \sin^2(k/2)$.
For values $|a| > 2$, $d_k$ does not possess any nodal point. 

Numerical simulations conducted for Eq.~(\ref{eq:lindbladian}) \footnote{The Lindbladian with quadratic Hermitian jump operator $\hat L$ satisfies a closed hierarchy \cite{hierarchy1,hierarchy2,hierarchy3}. In this case, the differential equation of two-point correlation functions is closed (see the Supplemental Material \cite{SM} for deriving the correlation dynamics). Therefore, we can numerically simulate the charge transport for system size up to $L=3000$.} distinctly reveal varied charge transport behaviors among the three cases.
Starting with $a=\sqrt 2$ and initiating the dynamics from a half-filling domain wall state $|\psi_o\rangle = |1,\dots,1,0,\dots,0\rangle$, the density evolution $\langle n_i\rangle_t$ demonstrates a ballistic front, as depicted in Fig.~\ref{fig:fig1}(a1). 
Evaluating the charge transport $C(t) = \sum_{i\ge 1} \langle n_i\rangle_t$, we observe a scaling behavior (after $t>1$): $\gamma C(t) \sim f(\gamma t)$, wherein the scaling function exhibits asymptotic behavior $f(x) \sim x^{2/3}$ as $x\rightarrow \infty$.
This behavior indicates a dynamical exponent converging to $z=3/2$ in the long-time regime \footnote{Note that the scaling function does not conform to the KPZ universality class due to a ballistic wavefront.}.
When fixing $a = 2$ [as shown in Fig.~\ref{fig:fig1}(b1)], the density evolution exhibits a superdiffusive front instead. 
After $t>1$, the transport converges to the form $\gamma C(t) \sim g(\gamma t)$ [displayed in Fig.~\ref{fig:fig1}(b2)] with a different scaling function $g(x) \sim x^{3/5}$ in the large $x$ limit, indicating a dynamical exponent of $z=5/3$.
For $a=3$, as demonstrated in Fig.~\ref{fig:fig1}(c1) [and in Fig.~\ref{fig:fig1}(c2) regarding the dynamical exponent], the transport displays apparent diffusive scaling in the long-time regime. 
This observation underscores the close relationship between the nodal structure of the dephasing quasi-particle and the dynamical scaling of the transport.

\textit{Wigner dynamics.}---
In Refs.~\cite{wigner1,wigner2}, the authors introduced a hydrodynamics framework tailored for free fermion systems characterized by quadratic jump operators.
The hydrodynamics of the free fermion system is captured by the Wigner distribution \cite{wigner3}: $n(x,k,t) \equiv \sum_s e^{iks} \langle \hat c^\dagger_{x+\frac{s}{2}} \hat c_{x-\frac{s}{2}} \rangle_t$.
This quantity essentially represents the particle density at position $x$ with momentum $k$ and offers a semiclassical perspective that accurately captures the system's dynamics in a coarse-grained sense. 
In the Supplemental Material \cite{SM}, we formally prove the exact Lindblad equation (\ref{eq:lindbladian}) leads to the following Wigner dynamics:
\begin{equation}\label{eq:ghd}
\begin{aligned}
	\partial_t n(x,k,t) =& -2\sin k \partial_x n(x,k,t)-\gamma |d_k|^2 n(x,k,t) \\
	&+\gamma|d_k|^2\int\frac{dq}{2\pi}|d_q|^2n(x,q,t).
\end{aligned}
\end{equation}
This equation describes a statistical process wherein a wave packet with momentum $k$ has a probability proportional to $|d_k|^2$ to shift to a different momentum.
The probability distribution of the new momentum $q$ follows the distribution $|d_q|^2$. 
We proceed to solve this linear equation employing the Green's function method: $n(x,k,t) = (G*n_o)(x,k,t) = \int G(y,k,t) n_o(x-y,k) dy$.
Taking the initial state as a domain wall configuration, i.e., $n_o(x,k) = \theta(-x)$, this expression simplifies to $n(x,k,t)= \int_x^\infty G(y,k,t) dy$, with the initial condition $G(x,k,0) = \delta(x)$.
The Green's function $G(x,k,t)$ can be efficiently simulated via a random walk approach \cite{wigner1} involving the following steps:
\begin{enumerate}
	\item The velocity is determined by momentum: $x'(t)=v[K(t)] = 2\sin K(t)$.
	\item The quantity $K(t)$ remains constant within each interval $[t_0,t_1), [t_1,t_2),\cdots$, with each interval being independent and following an exponential distribution with an average value of $\overline{t_{i+1}-t_i}= \gamma^{-1}|d_k|^{-2}$.
	\item The momenta $K_{i+1}$ are randomly distributed with a probability proportional to $p(k) \propto |d_k|^2$.
	\item The probability density $p(x,k,t)$ corresponds to the Green's function $G(x,k,t)$, which can be determined numerically by sampling various random trajectories.
\end{enumerate}
By employing this method and sampling multiple random trajectories, we obtain access to the scaling exponent in the long-time regime with high accuracy.

\begin{figure}[h]
	\centering
	\includegraphics[width=\linewidth]{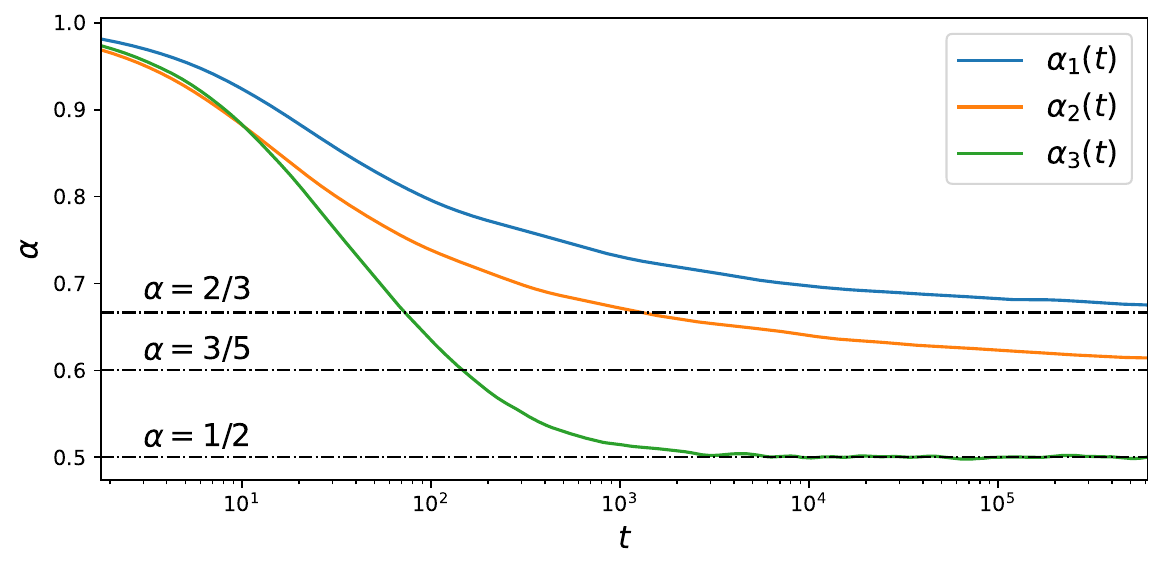}
	\caption{The dynamical exponent $\alpha(t) = d \log C(t)/d \log (t)$ of the charge transport for Wigner dynamics (\ref{eq:ghd}) with $\gamma=0.1$. These results are obtained from the random walk simulations with $5\times 10^6$ samples. The exponents $\alpha_1(t)$, $\alpha_2(t)$, and $\alpha_3(t)$ correspond to the cases $a=\sqrt{2}$, $a=2$, and $a=3$ respectively.}
	\label{fig:fig2}
\end{figure}

For the specific dephasing model involving quasi-particles in Eq.~(\ref{eq:generic-nodal}), Fig.~\ref{fig:fig1}(a2)(b2)(c2) showcase comparisons of charge transport between the exact Lindbladian dynamics on a 1D lattice and the Wigner dynamics.
Initially distinct, the Lindblad dynamics gradually converges to the Wigner dynamics beyond $t>1$. 
In Supplementary Material \cite{SM}, we demonstrate an agreement in the dynamics of density profiles obtained through both methods, particularly evident in the long-time regime. 
This agreement supports the accuracy of the hydrodynamics description. 
Leveraging this validation, we extend our numerical simulation using Wigner dynamics, pushing the simulation time to $t > 10^5$. 
As shown in Fig.~\ref{fig:fig2}, this extension enables a precise showcase of the convergence of the dynamical scaling $\alpha(t)$ towards $2/3$ and $3/5$.

\textit{L\'evy walk.}---
In the random walk picture, we show that the nodal point in the momentum distribution leads to the phenomenon of L\'{e}vy walk \cite{levy}.
In stark contrast to the Gaussian characteristics defining Brownian motion and standard diffusion, a L\'{e}vy walk constitutes a stochastic process dictated by a heavy-tailed probability distribution $p(l)$ governing step length $l$ of each transition. 
In the $l\rightarrow \infty$ limit, this distribution conforms to a power-law behavior: $p(l) \sim l^{-1-z}$, where $1<z<2$ is the L\'{e}vy exponent. 
When each step takes equal time, the cumulative displacement conforms to an asymptotic behavior: $X(t) = |\sum_i l_i| \sim t^z$. 
Hence, the dynamical exponent governing the system's behavior aligns with the L\'{e}vy exponent, confirming that a L\'{e}vy walk implies superdiffusion.

In systems where time-reversal and reflection symmetries are preserved, we define an indicator as $\nu = d_{0} d_{\pi}$, which indicates a nontrivial condition when $\nu \le 0$.
For quasi-particles in Eq.~(\ref{eq:generic-nodal}), this corresponds to the range $-2\le a \le 2$. 
A nontrivial $\nu < 0$ implies the existence of a nodal point at $k_o \in (0,\pi)$. 
We refer to this nodal point as a ``generic nodal point." By introducing $q=|k-k_o|$, in the vicinity of $q \approx 0$, the mean free path exhibits the asymptotic behavior: $l \sim \tau \sim |d_k|^{-2} \sim q^{-2}$. 
As $q$ approaches zero, the mean free path diverges.
A change of variable ($q \rightarrow l$) in the probabilistic distribution results in: $\int p(q)dq \sim \int q^{2} dq \sim \int l^{-1} d (l^{-1/2}) \sim \int l^{-5/2} dl$, leading to $p(l) \sim l^{-5/2}$. 
With a constant average time step: $\overline{t_{n+1}-t_n} = \int_k p(k) \tau_k = 1/\gamma$, this random walk behavior aligns with a L\'{e}vy walk, characterized by an exponent of $z=3/2$, consistent with our numerical simulations in Fig.~\ref{fig:fig1}(a2).

When $\nu=0$, $d_k$ possesses a nodal point at one of the high symmetry points, $k_o$, with a vanishing velocity $v_k \sim |k-k_o| = q$. 
The symmetry condition requires the dispersion of $d_k$ to be at least quadratic: $d_k \sim q^2$. 
As a result, the mean free path scales as $l \sim v\tau \sim q^{-3}$.
Then a change of variable ($q \rightarrow l$) leads to $\int p(q)dq \sim \int q^4 dq \sim \int l^{-4/3} d\left(l^{-1/3}\right) \sim \int l^{-8/3} dl$, which yields $p(l) \sim l^{-8/3}$.
The dynamical scaling exponent becomes $z = 5/3$, in accordance to Fig.~\ref{fig:fig1}(b2).

In cases where $\nu > 0$, there is typically no nodal point in $d_k$, resulting in a bounded mean free time: $\tau_k \le \tau_\text{max}$, and subsequently, a finite mean free path $l \le l_\text{max}$, resulting in ordinary diffusive behavior with $z = 2$, as shown in Fig.~\ref{fig:fig1}(c2).

\textit{Fine-tuning the dynamical exponents.}---
Beyond the symmetric setting, we can also leverage specific fine-tuned dephasing quasi-particles to attain higher-order dispersion near the nodal points. 
This diversity in dispersion yields various dynamical scaling behaviors.

Let us begin by considering a model with a single nodal point: $|d_k| \sim \sin^n \left[(k-k_o)/2\right]$.
This type of dispersion can be realized by selecting the following form for $\hat d_x$:
$
	\hat d_x = \mathcal{N}_n^{-1/2} \sum_{a=0}^n e^{-i a k_o} \hat c_{x+a},
$
where $\mathcal N_n$ is the normalization \footnote{The explicit form of the normalization is $$\mathcal N_n = \sum_{a=0}^n \begin{pmatrix} n \\ a \end{pmatrix}^2 = \frac{4^{n} \Gamma(n+\frac{1}{2})}{\sqrt{\pi } \Gamma(n)},$$ where $\Gamma(x)$ is the Gamma function.}.
In this case, $\hat d_x$ possesses a nodal point at $k_o$, exhibiting $n$-th order dispersion. 
Following similar derivations, the mean free path is given by $l \sim q^{2n}$, and the distribution takes the form:
$p(l) \sim l^{-1} \frac{d}{dl}(l^{-1/2n}) = l^{-1-(2n+1)/2n}$.
The dynamical scaling exponents that can be tuned in this scenario are given by $z_n = (2n+1)/(2n)$.

On the other hand, if we set $k_o=0$ (or equivalently $k_o=\pi$), the velocity $v_k \sim k$, resulting in $l \sim v \tau \sim k^{-(2n-1)}$. 
Consequently, the probability distribution becomes: $p(l) \sim l^{-1-(2n+1)/(2n-1)}$, which leads to the dynamical scaling exponent $z_n = (2n+1)/(2n-1)$.
Note that the derivation is valid only for $n \ge 2$. In the $n=1$ case, there would be usual diffusive transport.

\textit{Higher Dimensional Cases.}---
The analysis extends to higher dimensions, where nodal structures can also be nodal lines and surfaces.
To begin, the Wigner dynamics (Eq.~(\ref{eq:ghd})) naturally generalize to $D$-dimensions with slight modifications:
\begin{equation}
\begin{aligned}
	&\partial_t n(\bm x, \bm k, t) = 
	-2 \sum_{i=1}^D \sin(k_i) \partial_{x_i} n(\bm x,\bm k,t) \\
	&-\gamma |d_{\bm k}|^2 \left[n(\bm x,\bm k,t)-\int\frac{d^Dq}{(2\pi)^D} |d_{\bm q}|^2 n(\bm x,\bm q,t)\right].
\end{aligned}
\end{equation}
We refer to the supplemental Material \cite{SM} for the proof.
For simplicity, we assume the systems to be square/cubic lattices with nearest-neighbor hopping Hamiltonians, with dispersion relation $v_{\bm k} = 2 (\sin k_1 + \sin k_2)$ and $v_{\bm{k}} = 2 (\sin k_1 + \sin k_2 + \sin k_3)$ respectively.

For 2D systems, if the dephasing quasi-particle has the time-reversal and reflection symmetry, $d_{\bm k}$ is a real function on the Brillouin zone.
We can then similarly define three independent indicators: $\nu_1 = d_{(0,0)} d_{(\pi,0)}$, $\nu_2 = d_{(0,0)} d_{(0,\pi)}$, and $\nu_3 = d_{(0,0)} d_{(\pi,\pi)}$.
A negative value for these indicators indicates a nodal line $\bm k_o(\theta)$ in the Brillouin zone parametrized by $\theta \in [0,1)$. 
In proximity to this nodal line, we expect consistent behavior with $b_{\bm k} \sim k_{\perp}^{n}$, where $k_{\perp}$ represents the local variable orthogonal to the $\bm k_o$ curve.
The asymptotic probabilistic distribution $p(k_\perp)$ can be approximated by integrating out $k_\parallel$: $\int p(\bm k) dk_\parallel dk_\perp \sim \int k_\perp^{2n} d k_\perp$, i.e., $p(k_\perp)\sim k_\perp^{2n}$.
The mean free path is then $l\sim k_\perp^{-2n}$, indicating a 2D L\'{e}vy walk with a dynamical scaling exponent given by $z_n = (2n+1)/2n$.
The nodal line intersects with a high-symmetry point if any indicator yields zero.
The transport properties remain unchanged as they are determined by the segment of the curve with nonzero momentum.

When we relax the symmetry restriction on $\hat d_x$, $d_{\bm k}$ becomes a complex function on $k$.
Even in this case, nodal lines can exist without fine-tuning.
Consider the winding number $W[\mathcal C] = \oint_{\mathcal C} d_{\bm k}/\vert d_{\bm k}\vert$ of a contractible loop $\mathcal C$ in the Brillouin zone.
A nonzero winding number signifies the presence of a nodal point $k_o$ within the loop. 
For this analysis, we assume $p(\bm k) \sim |\bm k-\bm k_o|^{2n} \equiv |\bm q|^{2n}$ in the vicinity of the nodal point.
For a generic nodal point $k_o$ away from high-symmetry points, the mean free path is given by $l \sim \tau \sim |\bm q|^{-2n}$, and the probability distribution becomes: $\int p(\bm q) d\bm q^2 \sim \int q^{2n} q d q \sim \int l^{-2-1/n} dl$.
Consequently, the dynamical exponent is $z_n = (n+1)/n$.
If the nodal point is situated at one of the high-symmetry points, the mean free path becomes $l \sim v \tau \sim |\bm q|^{-(2n-1)}$, and the probability distribution can be expressed as $\int p(\bm q) d q^2 \sim \int  l^{-2-3/(2n-1)} dl$.
The dynamical exponent in this case is $z_n = (2n+2)/(2n-1)$.
For superdiffusion in this context, $n \ge 3$ is required; a smaller value of $n$ results in diffusive transport.

For the 3D systems, if time-reversal and reflection symmetry are present, we can similarly define seven indicators, which are the product of $\bm d_{0}$ and $\bm d_{\bm k}$ at one of seven high-symmetry points in the Brillouin zone. 
A negative sign in these indicators implies the presence of a nodal surface. 
Assuming that the dispersion near the nodal surface is proportional to the orthogonal component: $d_{\bm{k}} \sim k_\perp^n$, a similar calculation yields a dynamical exponent of $(2n+1)/(2n)$.
Without the symmetry constraint, we can similarly define winding number $W[\mathcal{C}]$ for a contractible loop $\mathcal{C}$; a non-zero winding implies a nodal line. 
Assuming the dispersion relation $d_{\bm{k}} \sim k_\perp^n$ near the curve, we obtain a dynamical exponent of $z_n = (n+1)/n$.

\textit{Conclusion.}---
This study uncovered a straightforward yet profound mechanism that leads to superdiffusive transport within noninteracting fermion systems subjected to local dephasing. 
Our findings demonstrate that we can fundamentally alter the system's behavior by extending the onsite particle dephasing to the dephasing of local quasi-particles featuring nodal points. 
The dynamics of a wave packet in this setting resemble the diffusive particle but with a unique feature: its mean free paths diverge when the momentum approaches the nodal point.

By studying the Wigner dynamics of the Lindbladian, we have rigorously mapped the system's behavior to that of a random walk. 
Notably, this random walk manifests as a L\'{e}vy walk, a well-established model of superdiffusion in physics. 
This mapping not only elucidates the physical underpinnings of the observed superdiffusion but also enables us to determine the dynamical exponent governing the system's behavior precisely. 
Furthermore, it empowers us to design and engineer models with different exact dynamical exponents, broadening our grasp of the phenomenon.
It is worth noting that this superdiffusive transport extends naturally to higher dimensions. 
This generality underscores the universality of the mechanism, offering valuable insights that can be applied across a spectrum of quantum many-body systems.

\begin{acknowledgements}
J.R. and Y.-P. W. thank Marko \v{Z}nidari\v{c} for his careful reading of the manuscript and for useful comments and suggestions. 
The numerical simulation of the Lindblad equation uses the \texttt{Julia} package \texttt{DifferentialEquation.jl} \cite{differentialequations}.
\end{acknowledgements}

\bibliography{ref}

\onecolumngrid

\newpage

\begin{center}
    \large
    \textbf{Supplemental Material for ``Superdiffusive Transport in Quasi-Particle Dephasing Models''}
    \normalsize
\end{center}

\section{Closed Hierarchy of the Correlation Function}

In this section, we will show that the dynamics governed by a Lindbladian consisting of free fermion Hamiltonian and Hermitian quadratic jump operators can be efficiently simulated due to a closed hierarchy \cite{hierarchy1,hierarchy2,hierarchy3} of the correlation function.
Specifically, the dynamics of the two-point correlation function can be formulated as a differential equation that is linear in itself and does not involve any multi-point correlations.

We first consider the Lindblad equation for operators:
\begin{equation}
	\partial_t \hat O = i [\hat H,\hat O] - \frac{\gamma}{2} \sum_n [\hat L_n[\hat L_n,\hat O]],
\end{equation}
where each jump operator is a Hermitian fermion bilinear:
\begin{equation}
	\hat L_x = \sum_{ab}d_{a}^* d_{b} c_{x+a}^\dagger c_{x+b} \equiv \sum_{ij}A_{x,ij}c_i^\dagger c_j,\quad A_{x,ij} = d^*_{i-x} d_{j-x}.
\end{equation}
Since we concern only the two-point correlation $G_{ij} = \langle c_i^\dagger c_j\rangle$, we can choose $\hat O_{ij} = c_i^\dagger c_j$.
Using the commutation relation $[c_i^\dagger c_j,c_k^\dagger c_l]=\delta_{jk}c_i^\dagger c_l -\delta_{il}c_k^\dagger c_j$, we know the following identity:
\begin{equation}\label{eq:quadratic-comm}
	\sum_{kl} [A_{kl}c_k^\dagger c_l, c_i^\dagger c_j]
	= \sum_k \left[ A_{ki} c_k^\dagger c_j - c_i^\dagger c_k A_{jk} \right].
\end{equation}
We can use the identity to calculate the commutator of two fermion bilinears and obtain the following:
\begin{equation}
	i \sum_{kl}H_{kl}[c_k^\dagger c_l, \hat O_{ij}]
	= i \sum_{kl} H_{kl} (\delta_{il}c_k^\dagger c_j -\delta_{jk}c_i^\dagger c_l)
	= i [H^T\cdot \hat O - \hat O\cdot H^T]_{ij}.
\end{equation}
Similarly, the double commutation in the second term is:
\begin{equation}
	- \frac{\gamma}{2} \sum_x [\hat L_x[\hat L_x,\hat O_{ij}]]
	= -\frac{\gamma}{2}\sum_x [(A^{*}_x)^2\cdot\hat O + \hat O\cdot (A_x^{*})^2 - 2 A_x^{*}\cdot \hat O \cdot A_x^{*}].
\end{equation}
Together, the EOM of the two-point correlation function is
\begin{equation}\label{eq:eom-2pt}
	\partial_t G = X^\dagger \cdot G + G \cdot X + \gamma \sum_x A_x^{*}\cdot G \cdot A_x^{*}\equiv \mathcal{L}[G],
\end{equation}
where $X = -i H^* - \frac{\gamma}{2}\sum_x (A_x^{*})^2$.

Note that the right-hand side of Eq.~(\ref{eq:eom-2pt}) is linear in $G$, the evolution of $G$ can be formally written as
\begin{equation}
	G(t) = e^{\mathcal L t}[G_0].
\end{equation}
To obtain the trajectory in the numerical simulation, simply implement the $\mathcal L[\cdot]$ action and insert the linear operator into a numerical solver for the differential equation.

\section{Wigner Dynamics for Quasi-Particle Dephasing Models}
\label{apx:ghd}

In this appendix, following Refs.~\cite{wigner1,wigner2}, we derive the Wigner dynamics of quasi-free Lindbiadian for general $d$-dimension.
The Hamiltonians are supposed to be the simplest free fermion model on the square lattice:
\begin{equation}
	\hat H = \sum_{\langle\bm x, \bm y\rangle} c_{\bm x}^\dagger c_{\bm y} + c_{\bm y}^\dagger c_{\bm x}.
\end{equation}
The Lindblad equation for operator $\hat O$ has the form:
\begin{equation}
	\partial_t \hat O = i [\hat H,\hat O] - \frac{\gamma}{2} \sum_n [\hat L_n[\hat L_n,\hat O]].
\end{equation}
We are considering the evolution of the operator 
\begin{equation}
	\hat n(\bm x,\bm k) \equiv \sum_s e^{i \bm k \cdot \bm s} c^\dagger_{\bm x+\frac{\bm s}{2}}c_{\bm x-\frac{\bm s}{2}},
\end{equation}
The hydrodynamics is then obtained by taking the expectation value: $n(\bm x,\bm k,t) = \langle \hat n(\bm x,\bm k)\rangle_t$.

\paragraph*{Hopping Hamiltonian}

We first consider the Hamiltonian part of the Lindbladian.
Using the identity
\begin{equation}
	\left[c_i^\dagger c_j,c_k^\dagger c_l\right]
	= c_i^\dagger[c_j,c_k^\dagger c_l] + [c_i^\dagger,c_k^\dagger c_l]c_j 
	=\delta_{jk}c_i^\dagger c_l -\delta_{il}c_k^\dagger c_j,
\end{equation}
we obtain the commutation relation
\begin{equation}
	[\hat H, c^\dagger_{\bm x} c_{\bm y}] = \sum_{i=1}^D \left( c^\dagger_{\bm x + \bm e_i} c_{\bm y} + c^\dagger_{\bm x - \bm e_i} c_{\bm y} -c^\dagger_{\bm x} c_{\bm y + \bm e_i} -c^\dagger_{\bm x} c_{\bm y - \bm e_i}\right),
\end{equation}
where $\bm e_i$ is the unit vector for each direction
and thus
\begin{equation}
\begin{aligned}
	i\left[\hat H,\hat n(\bm x,\bm k)\right] &= i\sum_{\bm s} e^{i\bm k\cdot \bm s} \sum_{i=1}^D\left[
	c^\dagger_{\bm x+\frac{\bm s}{2}+\bm e_i}c_{\bm x-\frac{\bm s}{2}}+
	c^\dagger_{\bm x+\frac{\bm s}{2}-\bm e_i}c_{\bm x-\frac{\bm s}{2}}-
	c^\dagger_{\bm x+\frac{\bm s}{2}}c_{\bm x-\frac{\bm s}{2}+\bm e_i}-
	c^\dagger_{\bm x+\frac{\bm s}{2}}c_{\bm x-\frac{\bm s}{2}-\bm e_i}\right] \\
	&= 2 \sum_{i=1}^D \sin(k_i) \left[\hat n\left(\bm x+\frac{\bm e_i}{2},\bm k\right)-\hat n\left(\bm x-\frac{\bm e_i}{2},\bm k\right)\right].
\end{aligned}
\end{equation}
In the coarse-grained 
\begin{equation}
	n\left(\bm x+\frac{\bm e_i}{2},\bm k,t\right)-n\left(\bm x-\frac{\bm e_i}{2},\bm k,t\right)
	\simeq \frac{\partial n}{\partial x_i} (\bm x,\bm k,t),
\end{equation}
So the Hamiltonian part of the hydrodynamics is
\begin{equation}
	d n(\bm x,\bm k,t) = -2 \sum_{i=1}^D \sin(k_i) \frac{\partial n}{\partial x_i}(\bm x,\bm k,t) dt.
\end{equation}

\paragraph*{Dissipation}
 Here, we consider the dephasing of the local quasi-particle at $y$, 
\begin{equation}
	\hat L_{y} = \sum_{a b} d_{a}^* d_{b} c^\dagger_{y+ a}c_{y+b}.
\end{equation}
We denote $A_{a b} = d_{a}^* d_{b}$, the commutator $[\hat L_{y}, \hat n(x, k)]$ is
\begin{equation}
\begin{aligned}
	\left[\hat L_{y}, \hat n(x,k)\right]
	&= \sum_{s,a b} e^{iks} A_{ab} \left[c^\dagger_{y+a} c_{y+b}, c^\dagger_{x+\frac{s}{2}} c_{x-\frac{s}{2}}\right] \\
	&= \sum_{s, a b} e^{iks} A_{ab} \left[
		\delta_{y-x+b,\frac{s}{2}} c^\dagger_{y+a}c_{x-\frac{s}{2}} -
		\delta_{x-y-a,\frac{s}{2}} c^\dagger_{x+\frac{s}{2}}c_{y+b}
		\right] \\
	&= \sum_{a b} A_{ab} \left[
		e^{-2ik(x-y-b)} c^\dagger_{y+a}c_{2x-y-b} -
		e^{+2ik(x-y-a)} c^\dagger_{2x-y-a}c_{y+b}
		\right].
\end{aligned}
\end{equation}
Using the fact 
\begin{equation}
	\frac{1}{N}\sum_p e^{-ipa} \hat n(x,p) = \frac{1}{N}\sum_p \sum_s e^{ip(s-a)} c_{x+\frac{s}{2}}^\dagger c_{x-\frac{s}{2}}
	= \sum_p \delta_{s,a} c_{x+\frac{s}{2}}^\dagger c_{x-\frac{s}{2}} = c_{x+\frac{a}{2}}^\dagger c_{x-\frac{a}{2}},
\end{equation}
the result is 
\begin{equation}
	\left[\hat L_y, \hat n(x,k)\right] = \frac{1}{N}\sum_{ab,p} A_{ab} e^{-ip(a-b)} \left[
		e^{-2i(k-p)(x-y-b)} \hat n\left(x+\frac{a-b}{2},p\right) -
		e^{+2i(k-p)(x-y-a)} \hat n\left(x-\frac{a-b}{2},p\right)
		\right].
\end{equation}
For the double commutator $\left[\hat L_y,\left[\hat L_y, \hat n(x,k)\right]\right]$, we need to replace the Wigner distribution in the right-hand side with $\left[\hat L_y, \hat n\right]$. 
There are four terms involved: 
\begin{equation*}
	\left[\hat L_y,\left[\hat L_y, \hat n(x,k)\right]\right]
	= S_1+S_2+S_3+S_4,
\end{equation*}
where $S_1$ and $S_2$ come from 
\begin{equation*}
	\hat n\left(x+\frac{a-b}{2},p\right)
	\ \longrightarrow \ 
	\left[\hat L_y,\hat n\left(x+\frac{a-b}{2},p\right)\right];
\end{equation*}
the $S_3$ and $S_4$ come from
\begin{equation*}
	\hat n\left(x-\frac{a-b}{2},p\right)
	\ \longrightarrow \ 
	\left[\hat L_y,\hat n\left(x-\frac{a-b}{2},p\right)\right].
\end{equation*}
In the following, we will simplify the expression term by term. 

For the first term $S_1$,
\begin{equation*}
\begin{aligned}
	S_1 &= \frac{1}{N^2}\sum_{abcd,pq,y} A_{ab}A_{cd}e^{-ip(a-b)-iq(c-d)}e^{-2i(k-p)(x-y-b)} 
	e^{-2i(p-q)(x+\frac{a-b}{2}-y-d)}
	\hat n\left(x+\frac{a-b+c-d}{2},q\right) \\
	&= \frac{1}{N}\sum_{abcd,pq} \left(\sum_y\frac{e^{2iy(k-q)}}{N}\right) A_{ab}A_{cd}e^{-ip(a-b)-iq(c-d)-2i(k-p)(x-b)} 
	e^{-2i(p-q)(x+\frac{a-b}{2}-d)}
	\hat n\left(x+\frac{a-b+c-d}{2},q\right) \\
	&= \frac{1}{N}\sum_{abcd,pq} \delta_{k,q} A_{ab}A_{cd}e^{-ip(a-b)-iq(c-d)-2i(k-p)(x-b)} 
	e^{-2i(p-q)(x+\frac{a-b}{2}-d)}
	\hat n\left(x+\frac{a-b+c-d}{2},q\right) \\
	&= \frac{1}{N}\sum_{abcd}\sum_{p} A_{ab}A_{cd}e^{-ip(a-b)-ik(c-d)} 
	e^{2i(k-p)(\frac{a+b}{2}-d)} \hat n\left(x+\frac{a-b+c-d}{2},k\right) \\
	&= \sum_{abcd} \left(\frac{1}{N}\sum_{p}e^{-2ip(a-d)}\right) A_{ab}A_{cd} 
	e^{ik(a+b-c-d)} \hat n\left(x+\frac{a-b+c-d}{2},k\right) \\
	&= \sum_{bc} (A^2)_{cb} e^{ik(b-c)} \hat n\left(x-\frac{b-c}{2},k\right).
\end{aligned}
\end{equation*}

The calculation for $S_4$ is similar to $S_1$:
\begin{equation*}
\begin{aligned}
	S_4 &= \frac{1}{N^2}\sum_{abcd,pq,y} A_{ab}A_{cd}e^{-ip(a-b)-iq(c-d)}e^{2i(k-p)(x-y-a)} 
	e^{2i(p-q)(x-\frac{a-b}{2}-y-c)}
	\hat n\left(x-\frac{a-b+c-d}{2},q\right) \\
	&= \frac{1}{N}\sum_{abcd,pq} \delta_{k,q} A_{ab}A_{cd}e^{-ip(a-b)-iq(c-d)}e^{2i(k-p)(x-a)} 
	e^{2i(p-q)(x-\frac{a-b}{2}-c)}
	\hat n\left(x-\frac{a-b+c-d}{2},q\right) \\
	&= \frac{1}{N}\sum_{abcd,p} A_{ab}A_{cd}e^{-ip(a-b)-ik(c-d)} 
	e^{i(p-k)(a+b-2c)} \hat n\left(x-\frac{a-b+c-d}{2},k\right) \\
	&= \sum_{abcd} \delta_{b,c} A_{ab}A_{cd} 
	e^{-ik(a+b-c-d)} \hat n\left(x-\frac{a-b+c-d}{2},k\right) \\
	&= \sum_{ad} (A^2)_{ad} e^{-ik(a-d)} \hat n\left(x-\frac{a-d}{2},k\right).
\end{aligned}
\end{equation*}
Since $\hat L_y$ is a particle number operator, $\hat L^2_y = \hat L_y$, i.e., $A^2 = A$.
Moreover, in the coarse-grained limit, we can approximate $\hat n(x-(b-c)/2,k)$ and $\hat n(x-(a-d)/2,k)$ with $\hat n(x,k)$.
Therefore,
\begin{equation}
	S_1 = S_4 = \sum_{a} d_a^* e^{-ika} \sum_b d_b e^{ikb} \hat n(x,k) = |d_k|^2 \hat n(x,k).
\end{equation}

Now we consider the $S_2$ part:
\begin{equation*}
\begin{aligned}
	S_2 &= -\frac{1}{N^2}\sum_{abcd,pq,y} A_{ab}A_{cd}e^{-ip(a-b)-iq(c-d)}e^{-2i(k-p)(x-y-b)} 
	e^{2i(p-q)(x+\frac{a-b}{2}-y-c)}
	\hat n\left(x+\frac{a-b-c+d}{2},q\right) \\
	&= -\frac{1}{N}\sum_{abcd,pq} \left(\sum_y \frac{e^{2iy(k-2p+q)}}{N}\right) A_{ab}A_{cd}e^{-ip(a-b)-iq(c-d)-2i(k-p)(x-b)} 
	e^{2i(p-q)(x+\frac{a-b}{2}-c)}
	\hat n\left(x+\frac{a-b-c+d}{2},q\right) \\
	&= -\frac{1}{N}\sum_{abcd,p} A_{ab}A_{cd}e^{-ip(a-b)-i(2p-k)(c-d)} 
	e^{i(k-p)(a+b-2c)} \hat n\left(x+\frac{a-b-c+d}{2},2p-k\right) \\
	&= -\sum_{abcd} \left(\sum_p \frac{e^{-2ip(a-d)}}{N}\right) A_{ab}A_{cd} 
	e^{ik(a+b-c-d)} \hat n\left(x+\frac{a-b-c+d}{2},2p-k\right) \\
	&= -\sum_{abcd} A_{ab}A_{cd} e^{ik(b-c)} \frac{1}{N}\sum_q e^{-iq(a-d)} \hat n\left(x+\frac{a-b-c+d}{2},q\right).
\end{aligned}
\end{equation*}
Using the coarse-graining approximation and replacing the momentum sum with the integral, we have:
\begin{equation}
\begin{aligned}
	S_2 &\simeq - \sum_{bc} d_b d_c^* e^{ik(b-c)}\int\frac{d^Dq}{(2\pi)^D} \sum_{ad}d_a^* d_d e^{-iq(a-d)} \hat n(x,q) \\
	&= - |d_k|^2 \int\frac{d^Dq}{(2\pi)^D} |d_q|^2 \hat n(x,q).
\end{aligned}
\end{equation}

Straightforward calculation shows $S_3 \simeq S_2$:
\begin{equation*}
\begin{aligned}
	S_3 &= -\frac{1}{N^2}\sum_{abcd,pq,y} A_{ab}A_{cd}e^{-ip(a-b)-iq(c-d)}e^{2i(k-p)(x-y-a)} 
	e^{-2i(p-q)(x+\frac{a-b}{2}-y-d)}
	\hat n\left(x-\frac{a-b-c+d}{2},q\right) \\
	&= -\frac{1}{N}\sum_{abcd,pq} \delta_{q,2p-k} A_{ab}A_{cd}e^{-ip(a-b)-iq(c-d)}e^{2i(k-p)(x-a)} 
	e^{-2i(p-q)(x-\frac{a-b}{2}-d)}
	\hat n\left(x-\frac{a-b-c+d}{2},q\right) \\
	&= -\sum_{abcd} A_{ab}A_{cd} 
	e^{-ik(a+b-c-d)} \int\frac{d^Dq}{(2\pi)^D} e^{2ip(b-c)} \hat n\left(x-\frac{a-b-c+d}{2},2p-k\right) \\
	&= -\sum_{abcd} A_{ab}A_{cd} e^{-ik(a-d)} \int\frac{d^Dq}{(2\pi)^D} e^{iq(b-c)}\hat n\left(x-\frac{a-b-c+d}{2},q\right) \\
	&\simeq - |d_k|^2 \int\frac{d^Dq}{(2\pi)^D} |d_q|^2 \hat n(x,q).
\end{aligned}
\end{equation*}
Therefore, we have proved that the hydrodynamics of the Wigner distribution is
\begin{equation}
	\partial_t n(\bm x, \bm k) = -2 \sum_{i=1}^D \sin(k_i) \partial_{x_i} n(\bm x,\bm k) -\gamma |d_{\bm k}|^2 n(\bm x,\bm k) +\gamma |d_{\bm k}|^2 \int\frac{d^Dq}{(2\pi)^D} |d_{\bm q}|^2 n(\bm x,\bm q).
\end{equation}

\section{Comparison between Exact Lindblad and Wigner Dynamics}

In this appendix, we compare the exact Lindblad dynamics
\begin{equation}\label{apx:lindblad}
	\partial_t \rho = -i[\hat H, \rho]-\frac{\gamma}{2}\sum_x [\hat d_x^\dagger \hat d_x,[\hat d_x^\dagger \hat d_x,\rho]]
\end{equation}
with the Wigner dynamics
\begin{equation}\label{apx:wigner}
	\partial_t n(x, k) = -2  \sin(k) \partial_{x} n(x, k) -\gamma |d_{k}|^2 n(x,k) +\gamma |d_{k}|^2 \int\frac{dq}{2\pi} |d_{q}|^2 n(x,q).
\end{equation}

\begin{figure}[H]
	\centering
	\includegraphics[width=\linewidth]{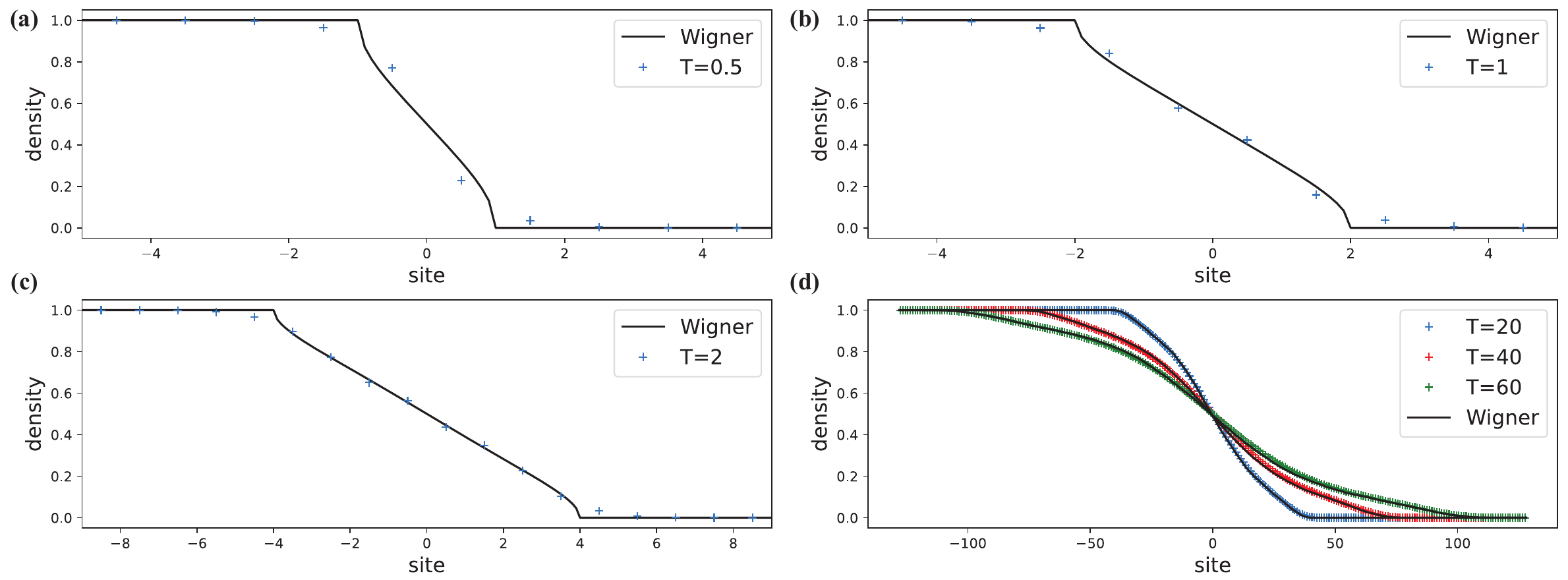}
	\caption{Comparison of the Lindblad equation (\ref{apx:lindblad}) and the Wigner dynamics (\ref{apx:wigner}) at $T=0.5, 1, 2, 20, 40, 60$, with $a=\sqrt 2$ and $\gamma=0.5$. The markers show the results from the Lindblad equation, and the solid line represents the results of Wigner dynamics.}
	\label{fig:f1}
\end{figure}

\begin{figure}[H]
	\centering
	\includegraphics[width=\linewidth]{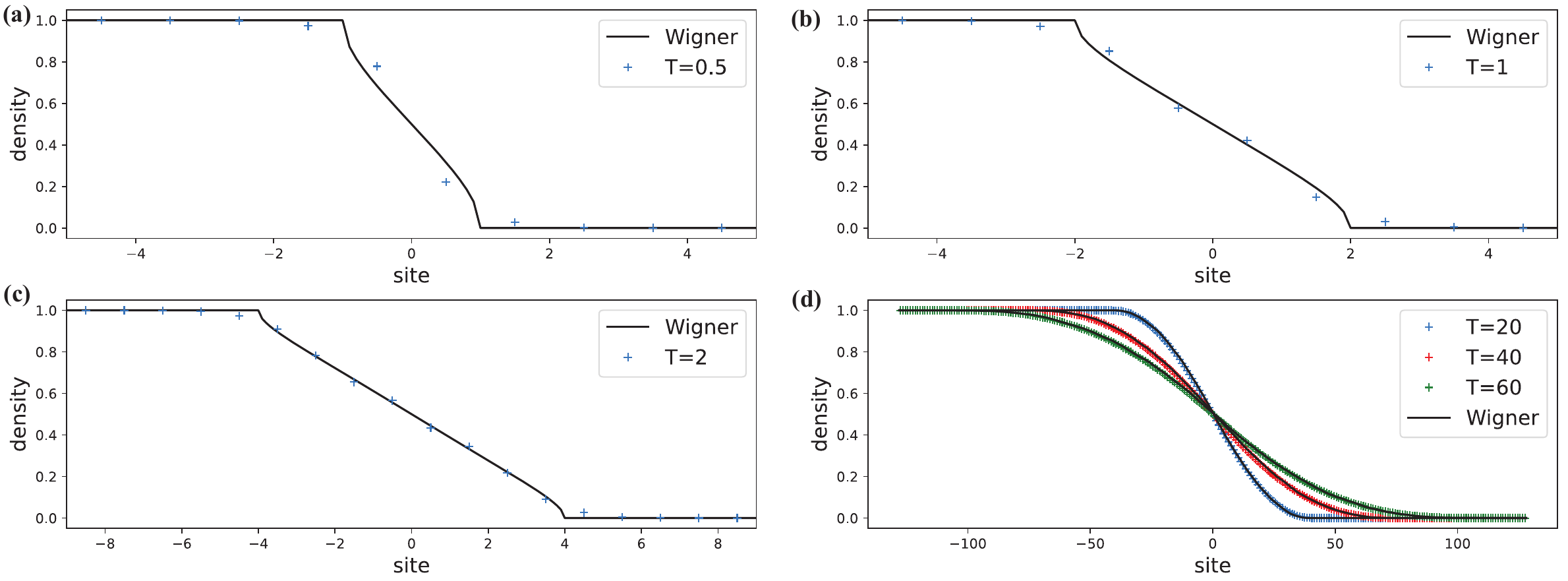}
	\caption{Comparison of the Lindblad equation (\ref{apx:lindblad}) and the Wigner dynamics (\ref{apx:wigner}) at $T=0.5, 1, 2, 20, 40, 60$, with $a=2$ and $\gamma=0.5$. The markers show the results from the Lindblad equation, and the solid line represents the results of Wigner dynamics.}
	\label{fig:f2}
\end{figure}

In terms of the dynamics of the density, we first notice that at a short time (e.g. $t=0.5$), [as displayed in Fig.~\ref{fig:f1}(a) for $a=\sqrt 2$ quasi-particle, and Fig.~\ref{fig:f2}(a) for $a=2$ quasi-particle] the exact Lindblad dynamics and the Wigner dynamics do not agree very well.
The comparison at later times ($t=1$ and $t=2$), as displayed in Fig.~\ref{fig:f1}(b)(c) for $a=\sqrt 2$ quasi-particle, and Fig.~\ref{fig:f2}(b)(c) for $a=2$ quasi-particle, numerics shows a better agreement.
When $t \ge 20$, as displayed in Fig.~\ref{fig:f1}(d) for $a=\sqrt 2$ quasi-particle, and Fig.~\ref{fig:f2}(d) for $a=2$ quasi-particle, we see a good agreement of two dynamics.

\end{document}